\documentclass[reprint, amsmath,amssymb,aps,prb]{revtex4-1}

\usepackage{graphicx}
\usepackage{dcolumn}
\usepackage{bm}
\usepackage[usenames]{color} 
\usepackage{amssymb} 
\usepackage{amsmath} 
\usepackage[utf8]{inputenc} 
\usepackage{braket}
\usepackage{mathptmx}

\newcommand{\bk}{\bm{k}}

\newcommand{\sro}{Sr$_2$RuO$_4$}

\newcommand{\kyotan} {\mathrm{i}}
\newcommand{\diff}{\mathrm{d}}
\newcommand{\imp}{\mathrm{imp}}
\newcommand{\isb}{\mathrm{ISB}}
\newcommand{\shc}{\sigma_{xy}^{\mathrm{S}}}

\begin{document}

\preprint{APS/123-QED}

\title{Controlling spin Hall effect by using a band anticrossing and nonmagnetic impurity scattering}


\author{T. Mizoguchi}
\email{mizoguchi@hosi.phys.s.u-tokyo.ac.jp}
\author{N. Arakawa}%
\affiliation{Department of Physics, The University of Tokyo, 
7-3-1 Hongo, Bunkyo-ku, Tokyo 113-0033, Japan 
}
\date{\today}

\begin{abstract}
The spin Hall effect (SHE) is one of the promising phenomena 
to utilize a spin current as spintronics devices, 
and the theoretical understanding of its microscopic mechanism is essential 
to know how to control its response. 
Although the SHE in multiorbital systems without inversion symmetry (IS) 
is expected to show several characteristic properties 
due to the cooperative roles of orbital degrees of freedom and a lack of IS, 
a theoretical understanding of the cooperative roles has been lacking. 
To clarify the cooperative roles, 
we study the spin Hall conductivity (SHC) derived 
by the linear-response theory 
for a $t_{2g}$-orbital tight-binding model of the $[001]$ surface or interface of Sr$_2$RuO$_4$ 
in the presence of dilute nonmagnetic impurities. 
We find that 
the band anticrossing, 
arising from a combination of orbital degrees of freedom and a lack of IS, 
causes an increase of magnitude
and a sign change of the SHC 
at some nonmagnetic impurity concentrations. 
Since a similar mechanism for controlling 
the magnitude and sign of the response of Hall effects 
works in other multiorbital systems without IS, 
our mechanism provides an ubiquitous method 
to control the magnitude and sign of the response of Hall effects 
in some multiorbital systems 
by introducing IS breaking and tuning of the nonmagnetic impurity concentration.

\begin{description}
\item[PACS numbers]
 72.25.Ba, 73.40.-c, 74.70.Pq
\end{description}
\end{abstract}

\pacs{Valid PACS appear here}
\maketitle
The spin Hall effect (SHE) has been intensively studied 
due to its applicability to spintronics devices and theoretical interests. 
In the SHE, an external electric field causes the spin current 
(i.e., the flow of the spin angular momentum)
which is perpendicular to this field~\cite{hirsch,dyakonov}. 
Since controlling a charge current is easier than controlling a spin current, 
the SHE has a great potential for spintronics devices~\cite{review-application}. 
In addition, it is crucial to understand the microscopic mechanism of the SHE 
since it gives us a deeper understanding of how to control its response, 
and this motivates further research. 

So far, we have partially understood the microscopic mechanisms of the SHE.
These are categorized into intrinsic and extrinsic mechanisms: 
The former is related to 
the electronic structure~\cite{murakami,sinova,QSHE,KM,kontani_tm}, 
while the latter is related to 
the multiple scattering of the doped impurities~\cite{hirsch,dyakonov,skew,sj,CreBru,pandey}. 
We emphasize that the intrinsic contribution arises from 
not only the Berry curvature term~\cite{murakami,sinova,QSHE,KM,onodanagaosa} 
(part of the Fermi sea term)
but also the other Fermi sea term and 
the Fermi surface term~\cite{streda,kontani_pt,kontani_tm,kontani_ahe}
which qualitatively differ from the Berry curvature term.
In several transition metals (TMs) and TM oxides,
the intrinsic mechanism is more important than 
the extrinsic mechanism 
because the intrinsic mechanism
often gives a large response~\cite{kontani_tm,kontani_pt}
and because the extrinsic mechanism is 
negligible for the weak scattering potential
of the doped nonmagnetic impurities.
Note that the realization of a such weak scattering potential is shown 
in a first-principles calculation in some cases~\cite{nakamura_ikeda}.\par
\begin{figure}[b]
\begin{center}
\includegraphics[width=8.5cm]{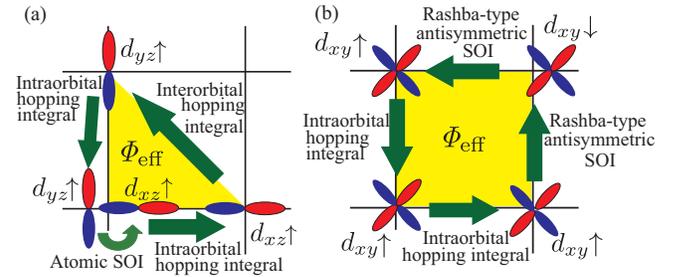}
\vspace{-10pt}
\caption{(Color online) Schematic pictures of some  processes
generating the effective flux $\varPhi_{\mathrm{eff}}$
in (a) the $t_{2g}$-orbital model with IS~\cite{kontani_tm} 
and (b) the single-orbital Rashba model without IS~\cite{sinova}.} 
\label{fig1}
\end{center}
\end{figure}

In general,
we can understand the origin of the intrinsic SHE 
by detecting 
how an electron acquires the 
Aharanov-Bohm-like phase~\cite{abkouka}
due to an effective flux.
For that detection,
it is helpful to consider the real-space motion of an electron 
whose first and final positions are the same 
because, by considering that motion,
we can discuss a phase factor of the wave function 
of an electron;
hereafter, we call such motion a process.
For example, 
in a multiorbital TM or TM oxide 
with inversion symmetry (IS)~\cite{pt_ex_1, pt_ex_2,kontani_pt,kontani_tm,tm_ex}, 
the spin-dependent effective flux is generated by the process
of using the $z$ component of the atomic spin-orbit interaction (SOI), 
the interorbital hopping integral between the orbitals connected by that $z$ component, 
and other hopping integrals [see Fig. \ref{fig1}(a)].  
In addition, we can understand
the SHE in a two-dimensional electron gas without IS~\cite{sinova,kato,2deg_ex}, 
whose electronic state is described by using the Rashba-type antisymmetric SOI~\cite{rashba} 
by considering the process
using the Rashba-type antisymmetric SOI and the intraorbital hopping integrals [see Fig. 1(b)]. 

Although there are many studies about the SHE 
in multiorbital systems with IS (e.g., Refs.~\onlinecite{kontani_pt,kontani_tm}) 
or single-orbital systems without IS (e.g., Refs.~\onlinecite{sinova,Rashba-CVC}), 
the characteristic properties of the SHE in a multiorbital system without IS are not well understood. 
In particular, 
the cooperative roles of the orbital degrees of freedom and lack of IS 
have not yet been clarified, 
although their combination will lead to several characteristic properties of the SHE.

To clarify these roles, 
it is necessary to study the SHE in a multiorbital system without IS 
by using a model considering  correctly both orbital degrees of freedom and IS breaking. 
It should be noted that since a combination of these leads to 
completely different results for several electronic properties 
from the results for a Rashba-type antisymmetric SOI,
the correct treatment beyond the Rashba-type antisymmetric SOI 
is significant for multiorbital systems. 
Actually, 
the momentum dependence of the $\bm{d}$ vector of a Cooper pair completely differs from 
that for a Rashba-type antisymmetric SOI 
due to the difference in the momentum dependence of the antisymmetric SOI; 
in the correct treatment, 
the antisymmetric SOI arises from 
the atomic SOI and the interorbital hopping integral 
due to local parity mixing, induced by IS breaking~\cite{yanase_sro,supplement}. 
Since the momentum dependence of the antisymmetric SOI is important even in discussing the SHE, 
a study about the SHE using the correct treatment is highly desirable. 

In this Rapid Communication, 
we study the SHE in a $t_{2g}$-orbital system without IS 
by using the correct treatment about 
the orbital degrees of freedom and a lack of IS 
beyond the Rashba-type antisymmetric SOI,
and reveal their cooperative roles in the SHE. 
In particular, 
we find that the band anticrossing due to the cooperative roles 
plays an important role in controlling 
the magnitude and sign of the spin Hall conductivity (SHC) of 
a multiorbital system without IS in the presence of dilute nonmagnetic impurities. 
After discussing the applicability of a similar mechanism, 
we propose a ubiquitous method to control 
the magnitude and sign of the Hall effects 
by using orbital degrees of freedom, 
IS breaking, and nonmagnetic impurity scattering.
\begin{figure}[t]
\hspace{-4pt}
\includegraphics[width=8.6cm]{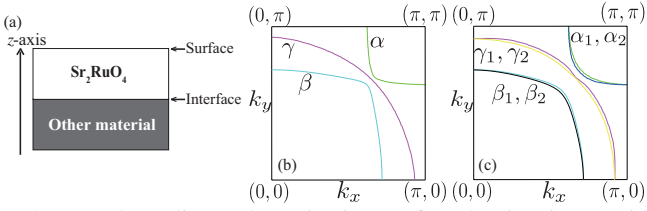}
\vspace{-12pt}
\caption{(Color online) Schematic picture of (a) the situation considered, 
and Fermi surfaces at (b) $t_{\isb}=0$ and (c) $t_{\isb}= 0.09$ eV.} 
\label{fig2}
\end{figure}

To discuss the SHE in a multiorbital system without IS, 
we consider a $t_{2g}$-orbital tight-binding model of 
the [001] surface or interface of Sr$_2$RuO$_4$ [Fig. \ref{fig2}(a)] 
in the presence of dilute nonmagnetic impurities: 
The Hamiltonian is $H=H_{0}+H_{\mathrm{LS}}+H_{\mathrm{ISB}}+H_{\mathrm{imp}}$, where 
$H_0  =
\sum_{\bk}
\sum_{a,b}
\sum_{\sigma} 
\epsilon_{ab} (\bk)c^{\dagger}_{\bk a\sigma} c_{\bk b\sigma}$, 
$H_{\mathrm{LS}} =
\sum_{\bk}
\sum_{a,b}
\sum_{\sigma,\sigma^{\prime}} 
\lambda[\bm{\ell}\cdot \bm{s}]_{a\sigma b\sigma^{\prime}}
c^{\dagger}_{\bk a\sigma} c_{\bk b\sigma^{\prime}}$, 
$H_{\mathrm{ISB}} =
\sum_{\bk}
\sum_{\sigma}
[V_x(\bk)c^{\dagger}_{\bk d_{yz}\sigma} c_{\bk d_{xy}\sigma} 
+ V_y(\bk) c^{\dagger}_{\bk d_{xz}\sigma} c_{\bk d_{xy}\sigma}+ \mathrm{H.c.}]$, 
and 
$H_{\mathrm{imp}} = 
\sum_{\bk,\bk^{\prime}}
\sum_{a}
\sum_{\sigma}
v_{\mathrm{imp}} 
\sum_{\bm{R}_j}e^{-\kyotan (\bk-\bk^{\prime}) \cdot \bm{R}_j} 
c^{\dagger}_{\bk a \sigma} c_{\bk^{\prime} a \sigma}$, 
with $a, b = d_{yz}, d_{xz}, d_{xy}$ and $\sigma, \sigma^{\prime} = \uparrow, \downarrow$. 
Hereafter, 
we set $c=\hbar=1$ and choose the lattice constant as unity.

$H_{0}$ and $H_{\mathrm{LS}}$ represent the kinetic energy and atomic SOI of Sr$_2$RuO$_4$, 
respectively; 
$\varepsilon_{ab}(\bk)$ are 
$\varepsilon_{d_{yz}d_{yz}} (\bk)= -2t_3 \cos k_x -2t_2 \cos k_y - \mu $, 
$\varepsilon_{d_{xz}d_{xz}}(\bk)= -2t_2 \cos k_x -2t_3\cos k_y -\mu $, 
$\varepsilon_{d_{xy}d_{xy}}(\bk) = -2t_1 (\cos k_x + \cos k_y) -4t_4 \cos k_x \cos k_y -\mu$, 
and $\varepsilon_{d_{xz}d_{yz}}(\bk)=\varepsilon_{d_{yz}d_{xz}}(\bk) =4t_5\sin k_x \sin k_y$; 
$[\bm{\ell}\cdot \bm{s}]_{a\sigma b\sigma^{\prime}}$ is 
$[\bm{\ell}\cdot \bm{s}]_{a\sigma b\sigma^{\prime}}=
-\frac{1}{2}(\delta_{a, d_{yz}}\delta_{b, d_{xy}}-\delta_{b, d_{xy}}\delta_{a, d_{yz}})
\delta_{\sigma, -\sigma^{\prime}}\textrm{sgn}(\sigma)
+\frac{i}{2}(\delta_{a, d_{xz}}\delta_{b, d_{xy}}-\delta_{b, d_{xy}}\delta_{a, d_{xz}})
\delta_{\sigma, -\sigma^{\prime}}\textrm{sgn}(\sigma)
+\frac{i}{2}(\delta_{a, d_{yz}}\delta_{b, d_{xz}}-\delta_{b, d_{xz}}\delta_{a, d_{yz}})
\delta_{\sigma, \sigma^{\prime}}\textrm{sgn}(\sigma)$, 
with $\textrm{sgn}(\sigma)=+1$ $(-1)$ for $\sigma=\uparrow$ ($\downarrow$). 
We choose these parameters so as to reproduce 
the experimentally observed Fermi surface of \sro~\cite{sro_dha}: 
$(t_1,t_2,t_3,t_4,t_5,\lambda)$ is $(0.45,0.675,0.09,0.18,0.03,0.045)$ (eV)~\cite{arakawa} 
and the chemical potential $\mu$ is determined so that 
the number of electrons per site is four. \par
$H_{\mathrm{ISB}}$ represents the interorbital hopping integral 
between the $d_{yz/xz}$ and $d_{xy}$ orbitals 
due to the local parity mixing
which is induced by 
IS breaking near the [001] surface or interface~\cite{yanase_sro,supplement}; 
$V_{x/y} (\bk)$ is $V_{x/y} (\bk)= 2\kyotan t_{\isb}\sin k_{x/y}$. 
(For details of the derivation,
see the Supplemental Material~\cite{supplement}.)
Note that
the second-order perturbation of $H_{\mathrm{LS}}$ and $H_{\mathrm{ISB}}$ gives 
a Rashba-type antisymmetric SOI 
when the orbital degeneracy is lifted by 
the large crystal-electric-field energy~\cite{yanase_sro}
(see the Supplemental Material~\cite{supplement}). 
In a case without IS, 
we set $t_{\isb}=0.09$ eV to make $t_{\isb}$ larger than $\lambda$ and $t_5$; 
as we will show, 
this condition is essential to obtain 
the magnitude and sign changes of the SHC. 
For comparison, 
we also consider the case at $t_{\isb}=0$ eV. 
As shown in the Supplemental Material~\cite{supplement},
to obtain finite intrinsic SHC,
only IS breaking is insufficient and 
finite atomic SOI is essential~\cite{murakami,kontani_tm}.

Diagonalizing $H_{\mathrm{band}}=\sum_{\bk}H_{\mathrm{band}}(\bk)=H_{0} + H_{\mathrm{LS}}+H_{\mathrm{ISB}}$, 
we obtain the band dispersions $E_n(\bk)$ and the unitary matrix $[U(\bk)]_{n\eta}$, 
where $\eta$ is $\eta \equiv (a,\sigma)$. 
Comparing the Fermi surfaces at $t_{\isb}=0$, and $0.09$ eV
shown in Figs. \ref{fig2}(b) and \ref{fig2}(c), respectively,
we see IS breaking causes 
the spin splitting of the $\alpha$, $\beta$, and $\gamma$ sheets. 
(At $t_{\isb}=0$ eV, 
the $\alpha$ and $\beta$ sheets are formed mainly by the $d_{xz}$ and $d_{yz}$ orbitals, 
and the $\gamma$ sheet is formed mainly by the $d_{xy}$ orbital~\cite{FS-orbital}.) 
In particular, 
the most drastic effect of IS breaking is 
the change of the curvature of the $\gamma$ sheet 
around $\bk\sim (\frac{2}{3} \pi,\frac{2}{3} \pi)$ 
due to the band anticrossing between the $d_{xy}$ and $d_{xz/yz}$ orbitals. 
In other words, 
this band anticrossing causes a change of the main orbital forming 
the $\gamma$ sheet around $\bk\sim (\frac{2}{3} \pi,\frac{2}{3} \pi)$ 
from the $d_{xy}$ orbital to the $d_{xz/yz}$ orbital. 
The necessary conditions for this band anticrossing are 
both 
that $t_{\isb}$ is larger than $\lambda$ and 
that several Fermi surface sheets 
whose main orbitals are connected by the transverse component of the atomic SOI
are close to each other 
in a certain area of the Brillouin zone. \par
$H_{\mathrm{imp}}$ represents the local scattering potential due to dilute nonmagnetic impurities. 
Assuming that the scattering is weak, 
we can treat its effects by the Born approximation; 
thus, $H_{\mathrm{imp}}$ affects 
the SHC through a self-energy correction 
and a current vertex correction due to 
the four-point vertex function~\cite{kontani_tm,kontani_pt,Rashba-CVC}. 
For simplicity, 
we assume that the main effects of $H_{\imp}$ arise from 
the band-independent quasiparticle damping $\gamma_{\imp}$
due to a self-energy correction, 
which is proportional to the nonmagnetic impurity concentration, $n_{\textrm{imp}}$. 
We have checked the neglected terms do not qualitatively change 
the results shown below~\cite{mizoguchi}.

\begin{figure*}[t]
\begin{center}
\includegraphics[width=13.8cm]{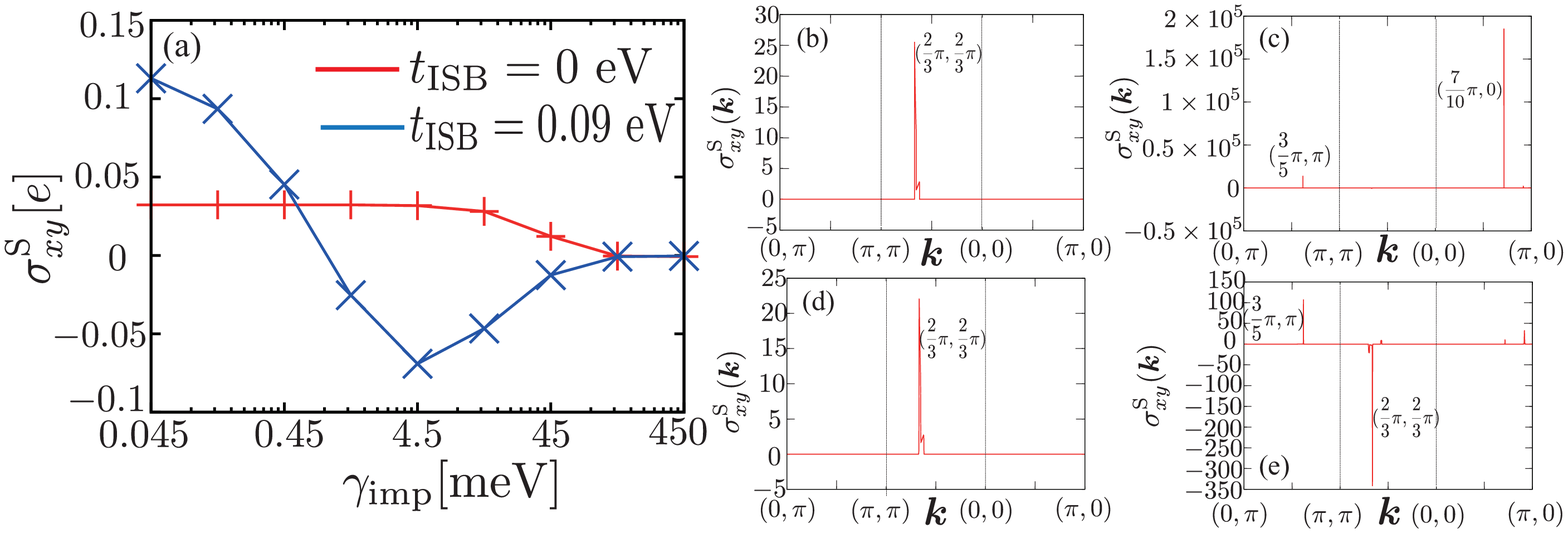}
\vspace{-10pt}
\caption{(Color online) 
(a) $\gamma_{\imp}$ dependence of $\shc$ at $t_{\isb}=0$ and $0.09$ eV
and $\bk$ dependence of $\shc(\bk)$ 
at (b) $(\gamma_{\imp}, t_{\isb})= 
(0.045\ \textrm{meV}, 0\ \textrm{eV})$, (c) $(0.045\ \textrm{meV}, 0.09\ \textrm{eV})$, 
(d) $(4.5\ \textrm{meV}, 0\ \textrm{eV})$, and (e) $(4.5\ \textrm{meV}, 0.09\ \textrm{eV})$.} 
\label{fig3}
\end{center}
\end{figure*}

Then, 
we derive the SHC by the linear-response theory~\cite{streda, kontani_pt}, 
$\sigma^{\mathrm{S}}_{xy}=\sigma^{\mathrm{S(I)}}_{xy}+\sigma^{\mathrm{S(I\hspace{-.1em}I)}}_{xy}$, 
where $\sigma^{\mathrm{S(I)}}_{xy}$ is the Fermi surface term,
\begin{align}
\sigma^{\mathrm{S(I)}}_{xy}=&\ 
\frac{1}{N}
\sum\limits_{\bm{k}} 
\sum\limits_{\{\eta_1\}} 
\int_{-\infty}^{\infty} \frac{\diff \omega}{2\pi}  
\Bigl(-\frac{\partial f(\omega)}{\partial \omega}\Bigr)  
[j^{\mathrm{S}}_x(\bk)]_{\eta_1\eta_2}\notag \\ 
&\times[j^{\mathrm{C}}_y(\bk)]_{\eta_3\eta_4}
[G^{\mathrm{R}}(\bm{k},\omega)]_{\eta_2\eta_3} [G^{\mathrm{A}}(\bk,\omega)]_{\eta_4\eta_1}, 
\label{Fermi surface}
\end{align}
and $\sigma^{\mathrm{S(I\hspace{-.1em}I)}}_{xy}$ is the Fermi sea term,
\begin{align}
\sigma^{\mathrm{S(I\hspace{-.1em}I)}}_{xy}=&\ 
\frac{1}{N}
\sum\limits_{\bm{k}} 
\sum\limits_{\{\eta_1\}} 
\int _{-\infty} ^{\infty} \frac{\diff \omega}{2\pi}  
f(\omega) 
\mathrm{Re}\Bigl\{ 
[j^{\mathrm{S}}_x(\bk)]_{\eta_1 \eta_2} 
[j^{\mathrm{C}}_y(\bk)]_{\eta_3 \eta_4}\notag \\
&\times 
([G^{\mathrm{R}}(\bm{k},\omega)]_{\eta_2 \eta_3} 
\tfrac{\overleftrightarrow{\partial}}{\partial \omega}
[G^{\mathrm{R}}(\bm{k},\omega)]_{\eta_4 \eta_1})
\Bigr\}. \label{Fermi sea}
\end{align}
Here 
$\sum_{\{\eta_1\}}$ 
is $\sum_{\{\eta_1\}}\equiv\sum_{\eta_1,\eta_2,\eta_3,\eta_4}$, 
$f(\omega)$ is the Fermi distribution function, $f(\omega)=\frac{1}{e^{\beta \omega}+1}$, 
$G^{\mathrm{R(A)}}(\bk,\omega)$ is the retarded (advanced) Green's function, 
$[G^{\mathrm{R(A)}}(\bk,\omega)]_{\eta_1\eta_2}= \sum_n [U^{\dagger}(\bk)]_{\eta_1n}
(\tfrac{1}{\omega - E_n(\bk) + (-) \kyotan \gamma_{\imp}}) [U(\bk)]_{n\eta_2} $, 
$j_y^{\mathrm{C}}(\bk)$ is the charge current operator, 
$j_y^{\mathrm{C}}(\bk) \equiv -e
\tfrac{\partial H_{\mathrm{band}}(\bk)}{\partial k_{y}}$, 
$j_x^{\mathrm{S}}(\bk)$ is the spin current operator, 
$j_x^{\mathrm{S}}(\bk) \equiv \frac{1}{2} 
[s_z\tfrac{\partial H_{\mathrm{band}}(\bk)}{\partial k_{x}}
+\tfrac{\partial H_{\mathrm{band}}(\bk)}{\partial k_{x}}s_z]$,  
and $(g\tfrac{\overleftrightarrow{\partial}}{\partial \omega}h)$ 
is $(g\tfrac{\overleftrightarrow{\partial}}{\partial \omega}h)\equiv 
g\tfrac{\partial h}{\partial \omega}-\tfrac{\partial g}{\partial \omega}h$. 
It should be noted, first, that 
in the clean limit, 
where $\gamma_{\imp}$ satisfies $\gamma_{\imp} \ll \Delta E(\bk)$ with 
$\Delta E(\bk)$ being the band splitting giving the dominant contribution to $\shc$, 
$\shc$ is independent of $\gamma_{\imp}$, 
and is given mainly by the Berry curvature term, part of $\sigma^{\mathrm{S(I\hspace{-.1em}I)}}_{xy}~\cite{kontani_ahe,kontani_pt}$;
second, that 
in the dirty limit, where $\gamma_{\imp}$ satisfies $\gamma_{\imp} \gg \Delta E(\bk)$, 
$\shc$ is proportional to $\gamma_{\imp}^{-3}$ 
and is given mainly by $\sigma^{\mathrm{S(I)}}_{xy}$~\cite{kontani_ahe,kontani_pt}.

We turn to results, 
where we use 
$20 \ 000 \times 20 \ 000$ meshes of the first Brillouin zone
because this size is necessary to suppress the finite-size effect in a clean region.
We first compare the $\gamma_{\imp}$ dependence of $\sigma_{xy}^{\mathrm{S}}$ 
at $t_{\isb}=0$ and $0.09$ eV in Fig. \ref{fig3}(a). 
Comparing the results at
$t_{\isb}=0$ eV and $0.09$ eV, we see three changes due to IS breaking: 
(i) an increase in a clean region ($\gamma_{\imp}<0.45$ meV), 
(ii) a sign change in a slightly dirty region ($0.45$ meV $\leq \gamma_{\imp} \leq 45$ meV), 
and (iii) the appearance of a minimum at $\gamma_{\imp}=4.5$ meV. 
These results suggest that 
the magnitude and sign of $\sigma_{xy}^{\mathrm{S}}$ can be controlled 
by using orbital degrees of freedom, IS breaking, and nonmagnetic impurity scattering. 
As shown in the Supplemental Material~\cite{supplement},
the above three changes hold
in the different 
values of $t_\isb$
if $t_\isb$ is larger than $\lambda$ 
and $t_5$.
Note, first, that 
in the range of $\gamma_{\imp}$ shown in Fig. \ref{fig3}(a), 
$\sigma^{\mathrm{S(I)}}_{xy}$ gives the main contribution to $\shc$
(see the Supplemental Material~\cite{supplement});
and, second, that the residual resistivity estimated by the Drude formula
is $0.1 \mu\Omega$cm in the clean region 
and 0.1 $-$ 10$\mu\Omega$cm in the slightly dirty region.
To clarify the origins of the above three changes, 
we analyze the $\bk$ dependence of $\sigma_{xy}^{\mathrm{S}}(\bk)$, 
defined as $\shc=\frac{1}{N}\sum_{\bk}\shc(\bk)$, at $t_{\isb}=0$ and $0.09$ eV. 
From the results at $\gamma_{\textrm{imp}}=0.045$ meV 
shown in Figs. \ref{fig3}(b) and \ref{fig3}(c), 
we see as $t_{\isb}$ changes from $0$ to $0.09$ eV, 
the main contribution to $\shc$ in the clean region 
changes from the region around $\bk\sim (\frac{2}{3} \pi,\frac{2}{3} \pi)$ 
to the region around $\bk\sim (\frac{7}{10} \pi,0)$. 
Since the latter main contribution is larger than the former one, 
change (i) arises from 
the evolution of the larger positive-sign contribution around $\bk\sim (\frac{7}{10} \pi,0)$. 
Then, 
from the results at $\gamma_{\textrm{imp}}=4.5$ meV 
shown in Figs. \ref{fig3}(d) and \ref{fig3}(e), 
we see that the change of $t_{\isb}$ from $0$ to $0.09$ eV 
in the slightly dirty region 
leads to the sign change of the main contribution to $\shc$ 
around $\bk\sim  (\frac{2}{3} \pi,\frac{2}{3} \pi)$ 
from positive to negative. 
Thus, 
this sign change is the origin of change (ii). 
Moreover, 
combining the results in the clean and the slightly dirty regions, 
we find 
change (iii) arises from the competition between 
the opposite-sign contributions around $\bk\sim  (\frac{7}{10} \pi,0)$ 
and $\bk\sim  (\frac{2}{3} \pi,\frac{2}{3} \pi)$. 

In addition, 
to understand how each $t_{2g}$ orbital contributes to 
each $\bk$ component of $\shc(\bk)$ at $t_{\isb}= 0$ and $0.09$ eV, 
we analyze orbital-decomposed SHCs, 
obtained by the equations that 
the summations with respect to orbital and spin indices 
in Eqs. (\ref{Fermi surface}) and (\ref{Fermi sea}) are restricted. 

\begin{figure*}[t]
\begin{center}
\includegraphics[width=15cm]{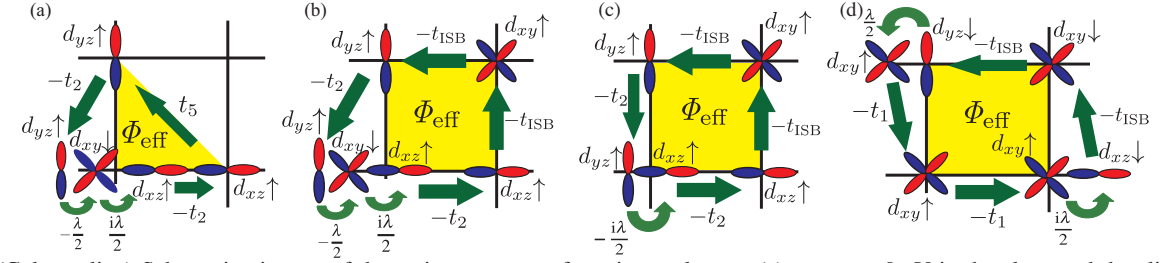}
\vspace{-14pt}
\caption{(Color online) 
Schematic pictures of the main  processes of a spin-up electron 
(a) at $t_{\isb}=0$ eV in the clean and the slightly dirty region, 
and at $t_{\isb}=0.09$ eV in (b) the clean and (c) the slightly dirty region, 
and of (d) an extended Rashba-type  process. 
} 
\label{fig4}
\end{center}
\end{figure*}

Before the results at $t_{\isb}=0.09$ eV, 
we explain the relation between the main $\bk$ component of $\shc(\bk)$ 
and each $t_{2g}$ orbital at $t_{\isb}=0$ eV. 
Considering all the possible orbital-decomposed SHCs which 
give the finite contribution to $\shc$ 
and calculating these values, 
we find that 
in the clean and the slightly dirty regions, 
the main contribution to $\shc$ arises from 
the term containing $[j^{\mathrm{S}}_x(\bk)]_{d_{xz}\sigma d_{xz}\sigma}
[j^{\mathrm{C}}_y(\bk)]_{d_{yz}\sigma d_{yz}\sigma}$ 
around $\bk\sim (\frac{2}{3}\pi, \frac{2}{3}\pi)$ near the Fermi level;
that main contribution arises from the process shown in Fig. \ref{fig4}(a). 
Although the process shown in Fig. \ref{fig1}(a) 
contributes to that term, 
this contribution is smaller than 
the above contribution [i.e., Fig. \ref{fig4}(a)] 
in the clean and the slightly dirty regions 
since the ratio of the magnitude of the former to the latter 
is roughly proportional to $\gamma_{\textrm{imp}}/\lambda$ [see Eqs. (6) and (7) in Ref.~\onlinecite{kontani_tm}]. 
Thus, all the $t_{2g}$ orbitals around $\bk\sim (\frac{2}{3}\pi, \frac{2}{3}\pi)$ 
near the Fermi level play an important role 
in the SHE at $t_{\isb}=0$ eV. 

We go on to explain the relation at $t_{\isb}=0.09$ eV. 
By using a similar analysis, 
we find that in the clean region, 
the main contribution to $\shc$ arises from 
the term containing $[j^{\mathrm{S}}_x(\bk)]_{d_{xz}\sigma d_{xz}\sigma}
[j^{\mathrm{C}}_y(\bk)]_{d_{xz}\sigma^{\prime} d_{xy}\sigma^{\prime}}$ 
or $[j^{\mathrm{S}}_x(\bk)]_{d_{xz}\sigma d_{xz}\sigma}
[j^{\mathrm{C}}_y(\bk)]_{d_{xy}\sigma^{\prime}d_{xz}\sigma^{\prime}}$ 
around $\bk\sim (\frac{7}{10}\pi, 0)$ near the Fermi level, 
whose main contribution arises from the process 
shown in Fig. \ref{fig4}(b). 
We also find that in the slightly dirty region, 
the main contribution to $\shc$ arises from the term containing 
$[j^{\mathrm{S}}_x(\bk)]_{d_{xz}\sigma d_{xz}\sigma}
[j^{\mathrm{C}}_y(\bk)]_{d_{yz}\sigma d_{yz}\sigma}$ 
around $\bk\sim (\frac{2}{3}\pi, \frac{2}{3}\pi)$ near the Fermi level, 
whose main contribution arises from the 
process shown in Fig. \ref{fig4}(c). 
We should note, first, that 
these processes can be regarded as not the extended Rashba-type ones, 
but the characteristic ones of a multiorbital system without IS 
since these completely differ from 
the extended Rashba-type process such as in Fig. \ref{fig4}(d); 
and, second, that 
in the slightly dirty region, 
although the contribution arising from the process shown in Fig. \ref{fig4}(a) 
is of opposite sign to the main contribution [i.e., Fig. \ref{fig4}(c)], 
the former is smaller due to $t_{\isb} > t_{5}$ and $t_{\isb} > \lambda$. 
Thus, all the $t_{2g}$ orbitals 
around $\bk\sim (\frac{7}{10}\pi, 0)$ and $(\frac{2}{3}\pi, \frac{2}{3}\pi)$,
which are 
affected by the spin splitting due to IS breaking, 
near the Fermi level 
become important in the SHE at $t_{\isb}=0.09$ eV. 
In particular, 
the sign change of the contribution around $\bk\sim (\frac{2}{3}\pi, \frac{2}{3}\pi)$ 
due to the band anticrossing 
and its competition with 
the opposite-sign contribution around $\bk\sim (\frac{7}{10}\pi, 0)$ 
are vital to obtain the magnitude and sign change of the SHC 
as a function of $\gamma_{\textrm{imp}}$.\par
We emphasize that 
our mechanism for controlling the magnitude and sign 
of the SHC is qualitatively different from the mechanism 
proposed in the single-orbital model ~\cite{rashba_dh_2}
with the Rashba- and Dresselhaus-type~\cite{dresselhaus}
antisymmetric SOIs. 
This is because our mechanism does not need
a bulk IS breaking which is necessary to obtain 
a Dresselhaus-type antisymmetric SOI.
\par
We now discuss the applicability of our mechanism 
for controlling the magnitude and sign of the SHC to other systems. 
If the following three conditions are satisfied, 
we can control the magnitude and sign of the SHC in a multiorbital system 
by introducing IS breaking and tuning $n_{\imp}$. 
These conditions are, first, that 
there are several (at least two) same-sign contributions from $\shc(\bk)$ at some momenta 
in a case without IS in the absence of band anticrossing; 
second, that 
the band anticrossing occurs at one of these momenta 
due to a combination of orbital degrees of freedom and IS breaking; 
and, third, that 
the value of the band splitting at the momentum where the band anticrossing occurs 
differs from the values of the band splittings giving the other contributions. 
If the first and second conditions are satisfied, 
we have the different-sign components of $\shc(\bk)$ at some momenta 
since the band anticrossing causes the sign change of $\shc(\bk)$ at one of these momenta, 
which is a result of the change of the main orbital of the Fermi surface sheet. 
In addition, if the third condition is satisfied, 
we can control the magnitude and sign of $\shc(\bk)$ 
by tuning the value of $\gamma_{\imp}$ through $n_{\imp}$ 
since an increase of $\gamma_{\imp}$ causes 
a larger decrease of the contribution of $\shc(\bk)$ arising from $\Delta E_{1}(\bk)$ 
than a decrease of the contribution of $\shc(\bk)$ 
arising from $\Delta E_{2}(\bk) > \Delta E_{1}(\bk)$,
;$\Delta E_{1}(\bk)$ [$\Delta E_{2}(\bk)$] is the band gap
at the momentum from which the negative (positive) contribution comes.
Since a large value of $t_{\isb}$ is necessary for the band anticrossing 
and an increase of the potential of the local parity mixing due to IS breaking enhances $t_{\isb}$, 
these three conditions can be realized even in other multiorbital systems 
by introducing IS breaking and tuning $n_{\imp}$. 
The surface of SrTiO$_3$~\cite{ueno_sto,sto_arpes} 
and the interface between SrTiO$_3$ and LaAlO$_3$~\cite{ohtomo_sto,sto_fp,yanase_nakamura} 
may be good candidates. 

Moreover, 
a similar mechanism is applicable to 
other Hall effects 
such as the anomalous Hall effect~\cite{kontani_ahe,KL},
since the band anticrossing due to a combination of orbital degrees of freedom 
and IS breaking drastically affects the main  process(es) generating the effective flux.\par
Thus, 
our mechanism gives a ubiquitous method 
to control the magnitude and sign of the response of Hall effects 
in multiorbital systems without IS.

In summary, 
we studied the SHE in the $t_{2g}$-orbital tight-binding model 
of the [001] surface or interface of Sr$_2$RuO$_4$ 
in the presence of dilute nonmagnetic impurities 
on the basis of the linear-response theory. 
We found that 
the SHC shows a magnitude increase and sign change 
as a function of $\gamma_{\imp}$ 
when the band anticrossing occurs at $\bk=(\frac{2}{3}\pi, \frac{2}{3}\pi)$ 
due to the cooperative roles of the orbital degrees of freedom and IS breaking,
and its contribution to the SHC becomes dominant compared with 
the contributions from $\shc (\bk)$ at the other momenta by increasing $\gamma_{\imp}$.
Since a similar situation can be realized in other systems 
by tuning the potential of the local parity mixing due to IS breaking and $n_{\imp}$, 
we propose that 
the magnitude and sign of the response of Hall effects 
in some multiorbital systems can be controlled by 
introducing IS breaking and tuning $n_{\imp}$. 

\begin{acknowledgements}
We would like to thank M. Ogata, S. Murakami, and T. Okamoto for fruitful discussions and comments.
\end{acknowledgements}
   

%
 \end{document}


\title{Supplemental Material for \lq \lq Controlling spin Hall effect by using a band anticrossing and nonmagnetic impurity scattering"}

\author{T. Mizoguchi}
\email{mizoguchi@hosi.phys.s.u-tokyo.ac.jp}
\author{N. Arakawa}%
\affiliation{Department of Physics, The University of Tokyo, 
7-3-1 Hongo, Bunkyo-ku, Tokyo 113-0033, Japan 
}
\date{\today}
\begin{abstract}
In this Supplemental Material,
we explain a microscopic derivation of $H_{\mathrm{ISB}}$,
derive the Rashba-type antisymmetric spin-orbit interaction from our model in the single-orbital limit,
highlight the essential role of the atomic spin-orbit interaction in the intrinsic spin Hall effect,
present the $\gamma_{\mathrm{imp}}$ dependence of $\sigma_{xy}^{\mathrm{S}}$ at different values of $t_{\mathrm{ISB}}$,
and show the role of the Fermi surface term and Fermi sea term in the range of $\gamma_{\mathrm{imp}}$ considered in the main text.
\begin{description}
\item[PACS numbers]
 72.25.Ba, 73.40.-c, 74.70.Pq
\end{description}
\end{abstract}

\pacs{Valid PACS appear here}
\maketitle

\section{Derivation of $H_{\isb}$}
\begin{figure}[b]
\begin{center}
\includegraphics[width=12cm]{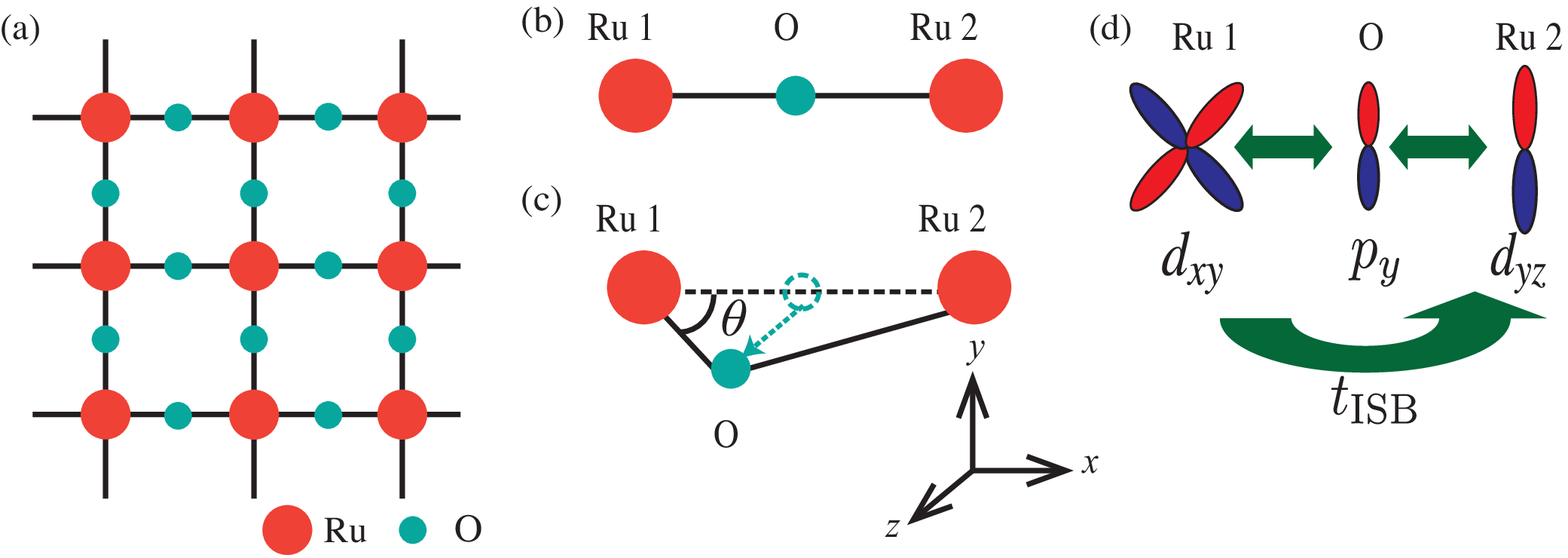}
\vspace{-10pt}
\caption{(Color online) Schematic pictures of (a) an $ab$ plane of Sr$_2$RuO$_4$,
the configuration of a Ru-O-Ru bond (b) with IS
and (c) without IS, 
and (d) an oxygen-mediated process which gives rise to $H_{\isb}$. 
} 
\label{figs1}
\end{center}
\end{figure}
In this section, 
we derive $H_{\isb}$ microscopically by using the Slater-Koster method~\cite{SK}
and show that $H_{\isb}$ originates from the local parity mixings
both between the $d_{yz}$ orbital at a Ru site and the $p_y$ orbital at an O site
along $x$ direction
and between the $d_{xz}$ orbital at a Ru site and the $p_x$ orbital at an O site
along $y$ direction.
Since the derivation of the term of $H_\isb$ along $y$ direction is similar way to that 
along $x$ direction, we explain only the derivation of the term along $x$ direction in the following.\par
Before deriving the term of $H_\isb$ along $x$ direction, we connect
the finite terms of the nearest-neighbor (n.n.) hopping integrals
of $H_0$ along $x$ direction with hopping integrals between the Ru-$t_{2g}$ orbitals
and O-$2p$ orbitals 
because the understanding of that connection is useful 
for the derivation of $H_\isb$.\par
Since Sr$_2$RuO$_4$ is a quasi-two-dimensional system~\cite{sro_dha},
and Ru sites in an $ab$ plane form a square lattice [see Fig. 1(a)],
the main n.n. hopping integrals for the Ru $t_{2g}$ orbitals along $x$ direction
come from the hopping integrals 
between the $d_{xz}$ and $p_z$ orbitals and 
between the $d_{xy}$ and $p_y$ orbitals:
the main hopping integral from Ru 1 to Ru 2 sites becomes  
 \begin{align}
 H_0^{d_2d_1} = & \frac{(H^{pd_2}_0)^\dagger H^{pd_1}_0 }{\Delta_{pd}}\notag \\
 = & 
\hat{d}^{\dagger}_{2}
\left(
\begin{array}{ccc}
0 & 0& 0\\
0 & -\frac{V_{pd\pi}^2}{\Delta_{pd}} &0\\
0 &0 & -\frac{V_{pd\pi}^2}{\Delta_{pd}} 
\end{array} 
\right)
\hat{d}_{1}, \label{dd}
\end{align}
where $\Delta_{pd}$ is the atomic energy difference between 
the Ru-$t_{2g}$ and O-$2p$ orbitals,
and $H_0^{pd_i}$ represents
the 
hopping integrals between the $2p$
orbitals at an O site 
and the $t_{2g}$ orbitals at a Ru $i$ sites 
in the configuration shown in Fig. \ref{figs1}(b),
i.e.,
\begin{subequations}
\begin{equation}
H^{pd_1}_0 = 
\hat{p}^{\dagger}
\left(
\begin{array}{ccc}
0 & 0& 0\\
0 & 0 & V_{pd\pi} \\
0 & V_{pd\pi} & 0
\end{array}
\right)
\hat{d}_1,  \label{pd}
\end{equation}
and
\begin{equation}
H^{pd_2}_0 = 
\hat{p}^{\dagger}
\left(
\begin{array}{ccc}
0 & 0& 0\\
0 & 0 & -V_{pd\pi} \\
0 & -V_{pd\pi} & 0
\end{array}
\right)
\hat{d}_2,  \label{pd2}
\end{equation}
\end{subequations}
with $\hat{p}^{\dagger} = (c^\dagger_{p_x} \ c^\dagger_{p_y} \ c^\dagger_{p_z})$,
$\hat{d}_i =(c^\dagger_{i,d_{yz}} \ c^\dagger_{i,d_{xz}} \ c^\dagger_{i,d_{xy}})^{\mathrm{T}}$,
and $V_{pd \pi}$, the Slater-Koster parameter~\cite{SK}
of $\pi$-bonding.
Since the matrix elements of $H_0^{d_2d_1}$ and 
$H_0^{d_1d_2}$ are the same,
$H_0$ becomes even with respect to momentum
[see $\varepsilon_{ab}(\bk)$ of $\hat{H_0}$ in the main text]. 
In Eq. (\ref{dd}),
we have neglected the orbital dependence 
of $\Delta_{pd}$ and $V_{pd\pi}$ for simplicity 
because the origin and expression of $H_\isb$ are the same except 
the expression of $t_\isb$ in terms of $\Delta_{pd}$ and $V_{pd\pi}$;
if we take account of orbital dependence realized in our model,
the expression of $t_\isb$ becomes more complex,
although the values of $t_{\isb}$ along $x$ and $y$ directions
remain the same,
resulting in the same expression of $H_{\isb}$.
In contrast to the expression of $H_\isb$,
we should consider such orbital dependence to obtain the correct
expression of $H_0$ of our model.
Since the aim of this section is microscopic derivation of $H_\isb$,
our simplification of $\Delta_{pd}$ and $V_{pd\pi}$ are valid.\par
However, if inversion symmetry (IS) is broken, the IS breaking may induce several hopping integrals
which are prohibited in the presence of IS. 
In the following, we consider such hopping integrals 
and show that these lead to $H_\isb$.
By creating a $[001]$ surface or an interface of Sr$_2$RuO$_4$,
the IS of $ab$ plane near the surface or the boundary 
is broken. 
That IS breaking causes the shifts of $O$ sites along
$z$ direction~\cite{onodanagaosa} due to the lack of one of the apical oxygens of a RuO$_6$ octahedral.
As a result of those shifts, 
we obtain the hopping integrals,
\begin{subequations}
\begin{equation}
\tilde{H}_{pd_1} =  
\hat{p}^{\dagger}
\left(
\begin{array}{ccc}
 0 & \sqrt{3}V_{pd\sigma} \sin \theta \cos ^2\theta+V_{pd\pi} \left(1-2 \cos ^2\theta\right) \sin \theta & 0 \\
 V_{pd\pi} \sin \theta & 0 & V_{pd\pi} \cos \theta \\
 0 & \sqrt{3}V_{pd\sigma} \cos \theta \sin ^2\theta+V_{pd\pi} \cos \theta \left(1-2 \sin ^2\theta\right) & 0 
\end{array}
\right)
\hat{d}_{1}, \label{pd_isb_p1}
\end{equation}
and 
\begin{equation}
\tilde{H}_{pd_2} =  
\hat{p}^{\dagger}
\left(
\begin{array}{ccc}
 0 & \sqrt{3}V_{pd\sigma} \sin \theta \cos ^2\theta+V_{pd\pi} \left(1-2 \cos ^2\theta \right) \sin \theta & 0 \\
 V_{pd\pi} \sin \theta & 0 & - V_{pd\pi} \cos \theta \\
 0 & - \sqrt{3}V_{pd\sigma} \cos \theta \sin ^2\theta- V_{pd\pi} \cos \theta \left(1-2 \sin ^2\theta \right) & 0 
\end{array}
\right)
\hat{d}_{2}, \label{pd_isb_p2}
\end{equation}
\end{subequations}
where $\theta$ is the angle 
between $x$ axis
and the bond between Ru 1 and O
[see Fig. \ref{figs1}(c)],
and $V_{pd\sigma}$ is the Slater-Koster parameter
of $\sigma$-bonding.
Here we neglect the change of $V_{pd\sigma}$, $V_{pd\pi}$,
and $\Delta_{pd}$ due to IS breaking for simplicity;
this treatment is sufficient for the purpose of this section 
because of the similar reason for neglecting the orbital dependence
of $\Delta_{pd}$ and $V_{pd\pi}$.
Then, to simplify Eqs. (\ref{pd_isb_p1}) and (\ref{pd_isb_p2}),
we assume that $\theta$ is sufficiently small 
(i.e., the shift is much smaller than the lattice constant).
As a result, we can use the approximation 
$\cos \theta \sim 1$, and $\sin \theta \sim \theta$,
and retain only the lowest order terms with respect to $\theta$.
Thus, we have the simple forms of the hopping integrals,
$\tilde{H}_{pd_i} = H_0^{pd_i} + H^\isb_{pd_i}$, 
with
\begin{equation}
 H^\isb_{pd_1} =  
\hat{p}^{\dagger}
\left(
\begin{array}{ccc}
 0 & \sqrt{3}V_{pd\sigma}  \theta - V_{pd\pi}  \theta & 0 \\
 V_{pd\pi}  \theta & 0 &0 \\
 0 &0  & 0 
\end{array}
\right)
\hat{d}_{1}, \label{pd_isb_3}
\end{equation}
and
\begin{equation}
 H^\isb_{pd_2} =  
\hat{p}^{\dagger}
\left(
\begin{array}{ccc}
 0 & \sqrt{3}V_{pd\sigma}  \theta - V_{pd\pi}  \theta & 0 \\
 V_{pd\pi} \theta & 0 & 0  \\
 0 & 0  & 0 
\end{array}
\right)
\hat{d}_{2}. \label{pd_isb_4}
\end{equation}
Equations (\ref{pd_isb_3}) and (\ref{pd_isb_4}) show that 
the IS breaking leads to the finite
hopping integral which is odd with respect to $z$ finite,
although in the presence of the IS, 
the finite hopping integrals
along $x$ direction should be even with respect to $z$.
This is the effect of the local parity mixing 
due to the IS breaking.
Thus, the new hopping integrals due to the IS breaking 
lead to an additional hopping 
of the $t_{2g}$ orbitals in combination with 
the hopping integrals existing in the presence of the IS breaking
[see Fig. \ref{figs1}(d)]:
that additional hopping integral either from Ru 1 and Ru 2
sites or from Ru 2 and Ru 1sites is, respectively,
\begin{align}
H_\isb^{d_2d_1} = & \frac{(H_0^{pd_2})^{\dagger} H_\isb^{pd_1} + ( H_\isb^{pd_2})^{\dagger}H_0^{pd_1}}{\Delta_{pd}} \notag \\
=& \hat{d}^{\dagger}_{2}
\left(
\begin{array}{ccc}
0 & 0 & -\frac{2 \theta V_{pd\pi}^2}{\Delta_{pd}}\\
0& 0 & 0 \\
\frac{2\theta V_{pd\pi}^2}{\Delta_{pd}} & 0 &0 
 \end{array}
\right)
\hat{d}_{1} \notag \\
=& \hat{d}^{\dagger}_{2}
\left(
\begin{array}{ccc}
0 & 0 & -t_\isb\\
0& 0 & 0 \\
t_\isb & 0 &0 
 \end{array}
\right)
\hat{d}_{1}, \label{dd1_isb}
\end{align}
or
\begin{align}
H_\isb^{d_1d_2} = & \frac{(H_0^{pd_1})^{\dagger} H_\isb^{pd_2} + ( H_\isb^{pd_1})^{\dagger}H_0^{pd_2}}{\Delta_{pd}} \notag \\
=& \hat{d}^{\dagger}_{1}
\left(
\begin{array}{ccc}
0 & 0 & \frac{2 \theta V_{pd\pi}^2}{\Delta_{pd}}\\
0& 0 & 0 \\
- \frac{2\theta V_{pd\pi}^2}{\Delta_{pd}} & 0 &0 
 \end{array}
\right)
\hat{d}_{2}\notag \\
=& \hat{d}^{\dagger}_{1}
\left(
\begin{array}{ccc}
0 & 0 &t_\isb\\
0& 0 & 0 \\
- t_\isb & 0 &0 
 \end{array}
\right)
\hat{d}_{2},
\end{align}
where $t_\isb$ is $t_\isb =\frac{2 \theta V_{pd\pi}^2}{\Delta_{pd}}$.
Since the matrix elements of $H^{d_2d_1}_\isb$ and $H_\isb^{d_1d_2}$ are the same 
except a factor (-1),
$H_\isb$ becomes odd with respect to momentum.
More precisely, the term of $H_\isb$ 
along $x$ direction,
$H_{\isb \parallel x}$, becomes 
\begin{align}
H_{\isb  \parallel x} =  & 
t_\isb \sum_{i}[ (c^{\dagger}_{\bm{R}_i,yz}c_{\bm{R}_i +\bm{e}_x,xy} + (c^{\dagger}_{\bm{R}_i,yz}c_{\bm{R}_i - \bm{e}_x,xy})+ (\mathrm{H.c.}) ] \notag \\
= & 2\mathrm{i}t_\isb\sum_{\bm{k}} \sin k_x (c^{\dagger}_{\bm{k}yz}c_{\bm{k}xy} -c^{\dagger}_{\bm{k}xy}c_{\bm{k}yz}),
\end{align}
with $\bm{e}_x$ being the unit lattice vector in $x$ direction.\par
We emphasize that we can straightforwardly extend 
the above derivation to case of other models.

\section{Derivation of the Rashba-type spin-orbit interaction}
\begin{figure}[b]
\begin{center}
\includegraphics[width=10cm]{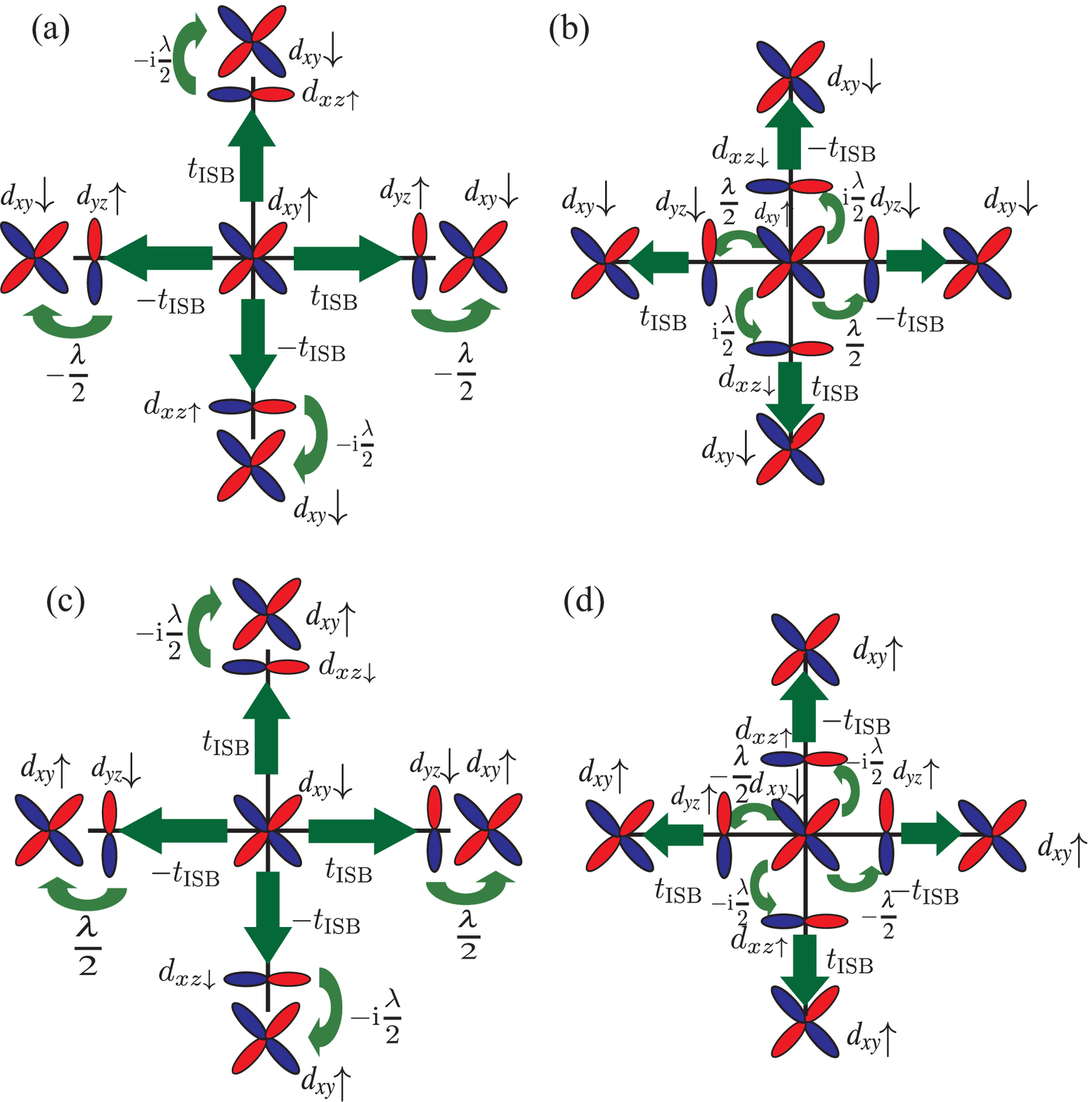}
\vspace{-10pt}
\caption{(Color online) Processes which give rise to the Rashba-type SOI.
Here we consider the second-order perturbation using the terms of $H_\isb$ and $H_\mathrm{LS}$. } 
\label{figs2}
\end{center}
\end{figure}

In this section,
we derive the Rashba-type antisymmetric spin-orbit interaction (SOI)~\cite{rashba} 
from our model by taking the single-orbital limit
and  
show that 
the Rashba-type antisymmetric SOI
originates from the combination 
of $H_\isb$, 
the transverse components of $H_{\mathrm{LS}}$,
and the large 
crystal field splitting.\par
To derive the Rashba-type antisymmetric SOI, 
we derive the antisymmetric SOI of our model in the single-orbital limit
because the Rashba-type antisymmetric SOI is an effective antisymmetric SOI 
for a single-orbital system~\cite{yanase_sro}.\par
To take the single-orbital limit of our model, 
we add $H_{\mathrm{CF}}$, 
\begin{equation}
H_{\mathrm{CF}} = - \Delta_{\mathrm{CF}}
\sum_{i,\sigma} c^{\dagger}_{i,d_{xy},\sigma}c_{i,d_{xy},\sigma},
\end{equation}
which represents the atomic energy difference 
between the $d_{xy}$ and $d_{xz/yz}$ orbitals,
to the Hamiltonian of our model, 
and choose $\Delta_{\mathrm{CF}}$ as being much larger than all the hopping integrals
and $\lambda$, and put $\mu$ in the range of $-\Delta_{\mathrm{CF}} < \mu <0 $.
Thus, in this limit, our model reduces to an effective single-orbital model
of the $d_{xy}$ orbital.
Because of a much larger value of $\Delta_{\mathrm{CF}}$, 
we can treat the effects of $H_\isb$ and $H_{\mathrm{LS}}$ 
as the perturbations against $H_{\mathrm{CF}}$.\par
Since the lowest-order terms are obtained 
by using the terms of $H_\isb$ and $H_{\mathrm{LS}}$ each once
[for all the possible processes see Figs. \ref{figs2}(a)-\ref{figs2}(d)],
we obtain the effective single-orbital antisymmetric SOI
in this single-orbital limit: 
\begin{align}
H_{\mathrm{ASSO}}
=  &  \frac{t_{\isb} \lambda}{\Delta_{\mathrm{CF}}}  [\sum_{i} - (c^{\dagger}_{\bm{R}_i+\bm{e}_x, d_{xy}\downarrow}c_{\bm{R}_i,d_{xy}\uparrow} 
-  c^{\dagger}_{\bm{R}_i- \bm{e}_x, d_{xy}\downarrow}c_{\bm{R}_i,d_{xy}\uparrow} ) \notag \\
 - & \mathrm{i} (c^{\dagger}_{\bm{R}_i+\bm{e}_y, d_{xy}\downarrow}c_{\bm{R}_i,d_{xy}\uparrow} - c^{\dagger}_{\bm{R}_i-\bm{e}_y, d_{xy}\downarrow}c_{\bm{R}_i,d_{xy}\uparrow}) 
+   (c^{\dagger}_{\bm{R}_i+\bm{e}_x, d_{xy}\uparrow}c_{\bm{R}_i,d_{xy}\downarrow} -c^{\dagger}_{\bm{R}_i- \bm{e}_x, d_{xy}\uparrow}c_{\bm{R}_i,d_{xy}\downarrow} )  \notag \\
-& \mathrm{i} (c^{\dagger}_{\bm{R}_i+\bm{e}_y, d_{xy}\uparrow}c_{\bm{R}_i,d_{xy}\downarrow} - c^{\dagger}_{\bm{R}_i-\bm{e}_y, d_{xy}\uparrow}c_{\bm{R}_i,d_{xy}\downarrow}) ] \notag \\
 = & \frac{2 t_{\isb}\lambda}{\Delta_{\mathrm{CF}}}\sum_{\bm{k}}
(c^{\dagger}_{\bm{k}d_{xy}\uparrow}, c^{\dagger}_{\bm{k}d_{xy}\downarrow})
\left(
\begin{array}{cc}
 0 & \sin k_y + i \sin k_x \\
 \sin k_y -  i \sin k_x & 0
\end{array}
\right)
\left(
\begin{array}{c}
c_{\bm{k}d_{xy}\uparrow}\\
c_{\bm{k}d_{xy}\downarrow}
 \end{array}
\right),\label{rashba_result}
\end{align}
where $\bm{e}_y$ is 
the unit lattice vector
in $y$ direction.
This equation shows that $H_{\mathrm{ASSO}}$ corresponds to the 
Rashba-type antisymmetric SOI with the coupling constant 
$\alpha_{\mathrm{Rashba}}=\frac{2t_\isb \lambda}{\Delta_{\mathrm{CF}}}$.
Thus, the Rashba-type antisymmetric SOI can be derived from our model
in the single-orbital limit by using the parity-mixing hopping and the transverse component of the atomic SOI.\par
It should be noted that if we do not take the single-orbital limit
(i.e., we consider a multiorbital system),
we cannot treat the effects of the $H_\isb$ and $H_{\mathrm{LS}}$ as the perturbations,
and those effects cannot be described by the simple Rashba-type antisymmetric SOI
(even by the sum of the Rashba-type antisymmetric SOI of each orbital).
Actually, as we pointed out in the main text,
the effects beyond the simple Rashba-type antisymmetric SOI
exist in a multiorbital system~\cite{yanase_sro}.\par
Thus, our formalism can describe not only an effective single-orbital system
with the Rashba-type antisymmetric SOI but also a multiorbital system
with a more complex antisymmetric SOI.
In other words,
our formalism has a potential for capturing the effects of the combination
of orbital degrees of freedom and IS breaking
beyond the Rashba-type SOI.
Actually, we highlight the importance of such effects
in the intrinsic spin Hall effect (SHE) (see the main text).

 \section{Essential role of the atomic SOI in the intrisic SHE}
\begin{figure*}[b]
\begin{center}
\includegraphics[width=8cm]{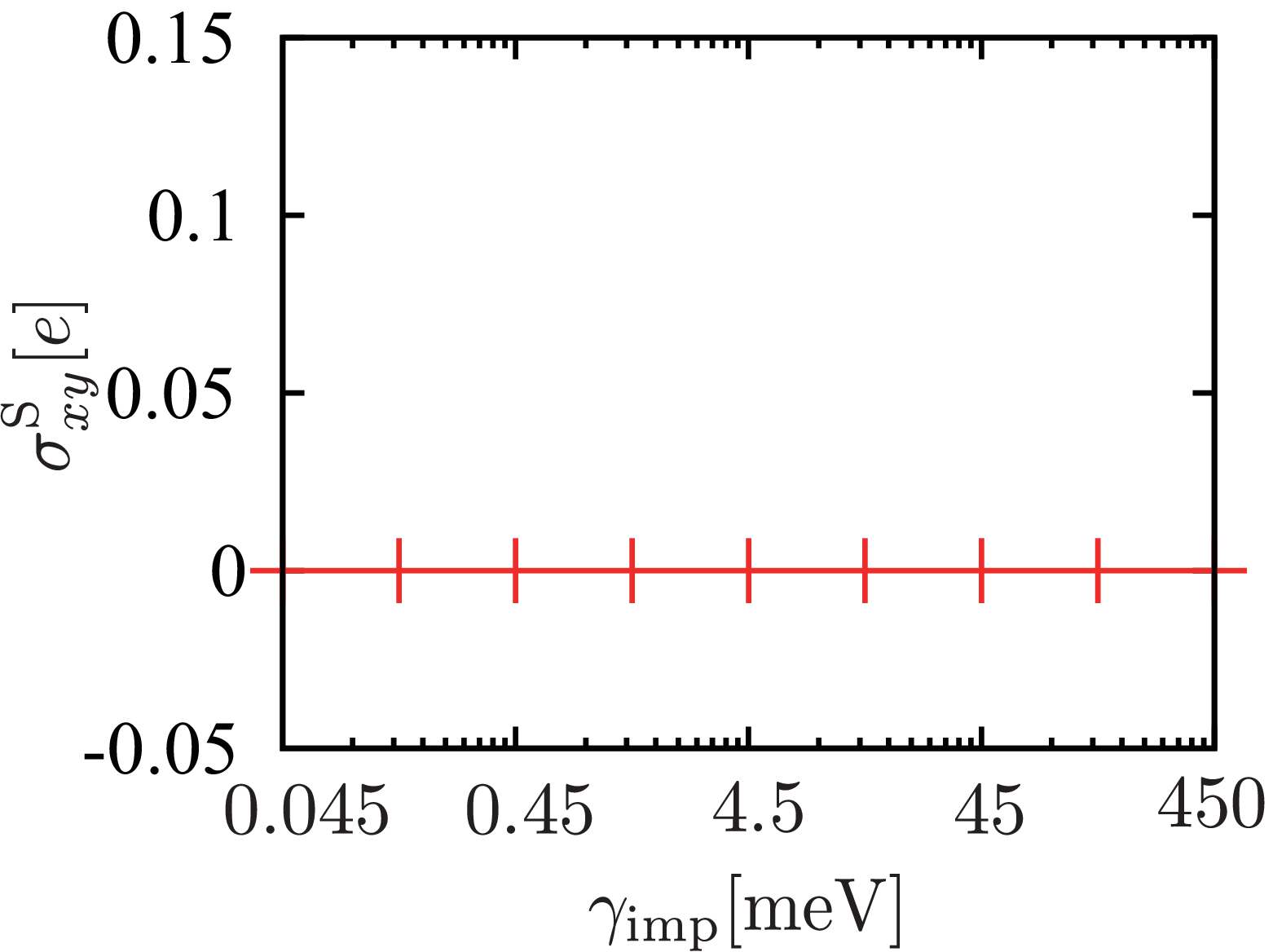}
\vspace{-10pt}
\caption{(Color online) 
$\gamma_\imp$ dependence of $\shc$ at $\lambda =0$ and $t_\isb=0.09$ eV.} 
\label{figs3}
\end{center}
\end{figure*}

In this section,
we present the $\gamma_\imp$ dependence of $\sigma_{xy}^{\mathrm{S}}$
at $\lambda=0$ and $t_\isb>0$ and highlight the essential role of the atomic SOI
in obtaining finite $\shc$.\par
Since the essential role of the atomic SOI in the intrinsic SHE 
has been confirmed in a multiorbital system with IS~\cite{kontani_tm},
we need to check the role of the atomic SOI in case without IS.\par
To check that role, we set $\lambda=0$ with keeping $t_\isb$ finite
and calculate $\sigma_{xy}^{\mathrm{S}}$ at several values of $\gamma_\imp$ in the same way
as the numerical calculation used in the main text.
The calculated result presented in Fig. \ref{figs3}
shows that only the IS breaking is insufficient 
to obtain the finite response of the SHE.
[As described in the main text
as well as the first section of this Supplemental Material,
the effects of the IS breaking are described by the terms of $H_\isb$.]
Combining that calculated result with the results presented in Fig. 3
of the main text, 
we deduce that the atomic SOI is essential to obtain finite response
of the intrinsic SHE even in case without IS.\par
Then, we can provide the brief explanation of the above result of the numerical calculation.
If we set $\lambda=0$, the Hamiltonian is 
decomposed into spin-up and spin-down sectors
and their matrix elements are completely the same.
In this case, the matrix elements of the spin current for spin-up sector
are the same as those for spin-down sector except a factor of $(-1)$,
and the matrix elements of the charge current and the Green's function
for spin-up sector are completely the same as the corresponding matrix elements
for spin-down sector. 
Thus, the contributions to $\shc$ from spin-up and spin-down sectors 
are exactly cancelled out.
Due to this exact cancellation,
$\shc$ becomes zero for $\lambda=0$ even without the IS.

\section{$\gamma_\imp$ dependence of $\shc$ at the different values of $t_\isb$}
\begin{figure}[t]
\begin{center}
\includegraphics[width=10cm]{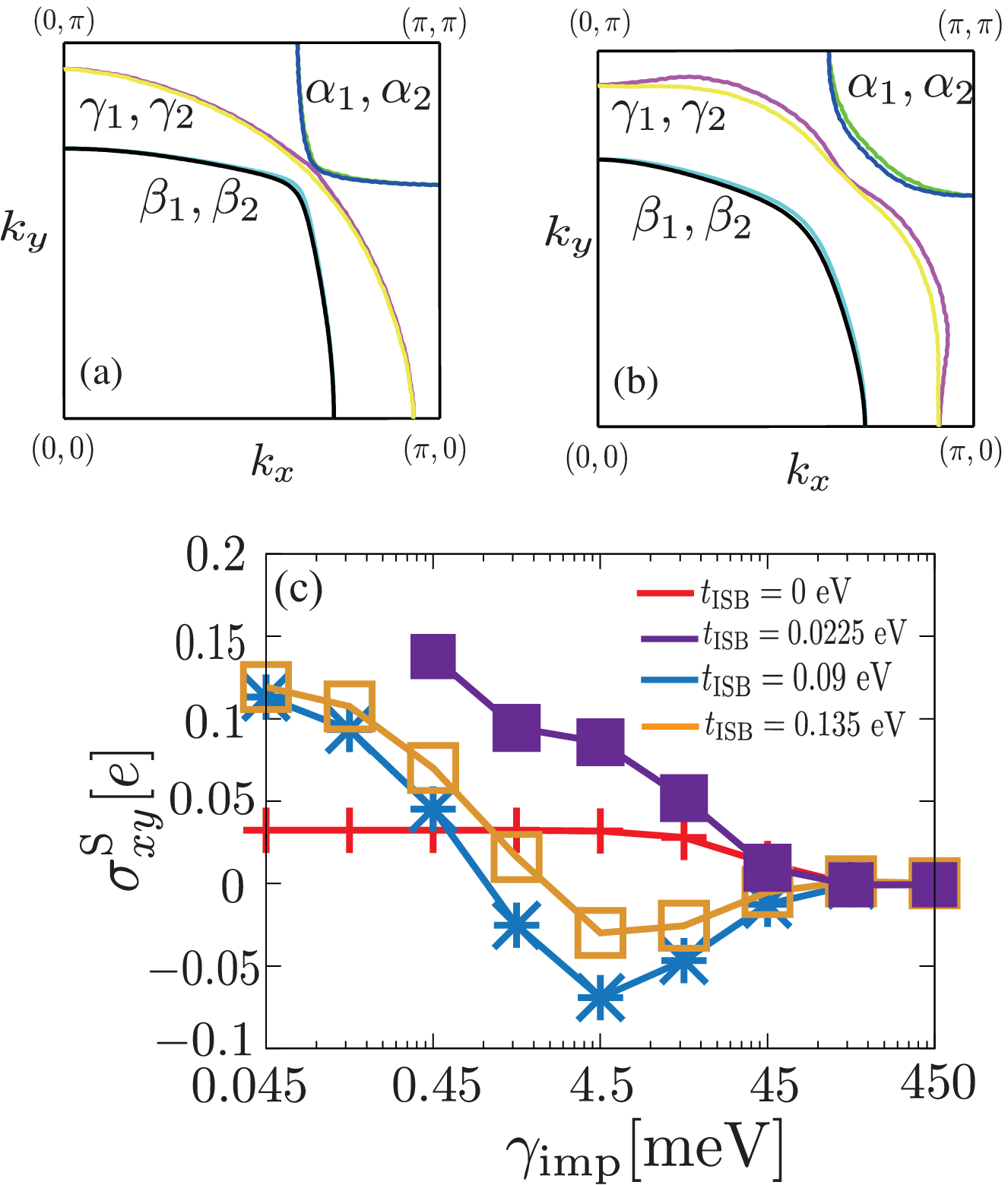}
\vspace{-10pt}
\caption{(Color online) 
The Fermi surface of (a) $t_\isb =0.0225$ eV and 
(b) $t_\isb =0.135$ eV, and (c) $\gamma_\imp$ dependence of $\sigma_{xy}^{\mathrm{S}}$ 
at $t_{\isb}=0$, 0.0225, 0.09, and 0.135 eV.} 
\label{figs4}
\end{center}
\end{figure}

In this section, we present the results of 
$\shc$ which support our statement about the importance of the band anticrossing
in the sign and magnitude change of $\shc$. \par
For a deeper understanding of the role of the band anticrossing due to the IS breaking,
we consider two other cases, case at $t_\isb=0.025$ eV and case at $t_\isb=0.135$ eV,
because the band anticrossing occurs only in the latter case
due to a large value of $t_\isb$ than the values of $t_5$ and $\lambda$. 
Actually, we see from Figs. \ref{figs4}(a) and \ref{figs4}(b)
that the curvature of $\gamma$ sheet around 
$\bm{k}=(\frac{2\pi}{3},\frac{2\pi}{3})$
does not change at $t_\isb=0.0225$ eV 
and changes at $t_\isb=0.135$ eV.
The change at $t_\isb=0.135$ eV is qualitatively the same as that at 
$t_\isb=0.09$ eV
[see Fig. 2(c) of the main text].\par
Then, we calculate the $\gamma_\imp$ dependence of $\shc$ at 
$t_\isb=0.0225$ eV and 0.135 eV
(where the other parameters are the same as case at $t_\isb=0.09$ eV)
and compare these results with the results at $t_\isb=0$ and 0.09 eV
in Fig. \ref{figs4}(c).
Figure \ref{figs4}(c) shows that although the magnitude increases
of $\shc$ occur at $t_\isb=0.0225$ eV and $t_\isb=0.135$ eV,
the sign changes of $\shc$ occur only at $t_\isb=0.135$ eV.
Those results indicate that we obtain not only the magnitude change of $\shc$
but also the sign change of $\shc$ only in the presence 
of the band anticrossing due to the IS breaking.
Thus, these results support the validity of our mechanism for 
controlling the magnitude and sign of $\shc$ by using the band anticrossing
and nonmagnetic impurity scattering.

\begin{figure}[b]
\begin{center}
\includegraphics[width=14cm]{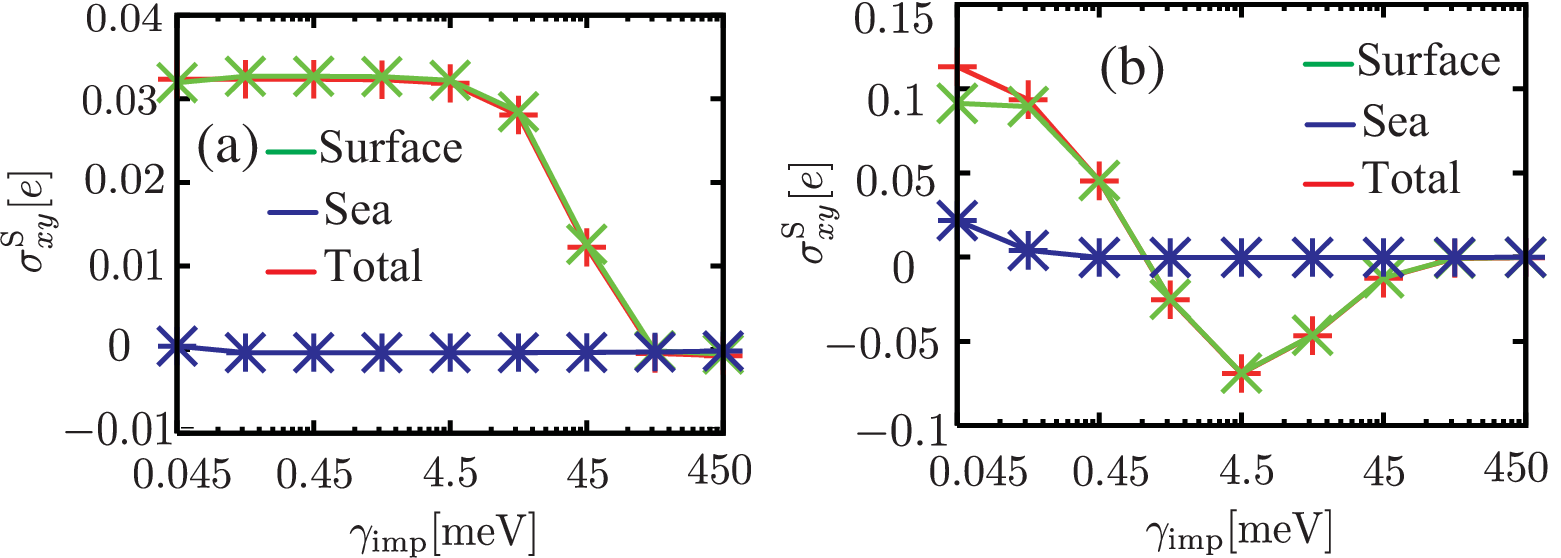}
\vspace{-10pt}
\caption{(Color online) $\gamma_\imp$ dependence of the Fermi surface term, $\sigma_{xy}^{\mathrm{S(I)}}$,
 and the Fermi sea term, $\sigma^{\mathrm{S(I\hspace{-.1em}I)}}_{xy}$,
at (a) $t_\isb=0$ eV and (b) $t_\isb=0.09$ eV.
Green lines represent $\sigma_{xy}^{\mathrm{S(I)}}$,
blue lines represent $\sigma^{\mathrm{S(I\hspace{-.1em}I)}}_{xy}$,
and red lines represent $\sigma_{xy}^{\mathrm{S}}=\sigma_{xy}^{\mathrm{S(I)}}+\sigma^{\mathrm{S(I\hspace{-.1em}I)}}_{xy}$.
} 
\label{figs5}
\end{center}
\end{figure}
\section{Dominance of the Fermi surface term of $\shc$}
In this section, 
we discuss the role of the Fermi surface term 
and the Fermi sea term 
in determining $\shc$ as a function of $\gamma_\imp$
at $t_\isb=0$ and 0.09 eV and show the dominance of the Fermi surface term 
in the parameter region we consider. \par
To discuss the role of the Fermi surface term
and the Fermi sea term,
we compare $\sigma_{xy}^{\mathrm{S(I)}}$
and $\sigma^{\mathrm{S(I\hspace{-.1em}I)}}_{xy}$
[determined by Eqs. (1) and (2) in the main text, respectively]
with $\sigma_{xy}^{\mathrm{S}}$
at several values of $\gamma_\imp$
in Figs. \ref{figs5}(a) and \ref{figs5}(b).
Those figures show that $\sigma^{\mathrm{S(I\hspace{-.1em}I)}}_{xy}$
is less dominant and that $\sigma_{xy}^{\mathrm{S(I)}}$
gives the dominant contribution to $\shc$.\par
Thus, the Fermi surface term 
(which cannot be described by the Berry curvature term,
as explained in the main text)
is dominant in determining $\shc$
in case not only with IS but also without IS
in the range of $0.045 \leq \gamma_\imp \leq 450$ (meV).